\documentclass[fleqn,usenatbib,letters]{mnras}

\usepackage{newtxtext,newtxmath}
\usepackage[T1]{fontenc}

\DeclareRobustCommand{\VAN}[3]{#2}
\let\VANthebibliography\thebibliography
\def\thebibliography{\DeclareRobustCommand{\VAN}[3]{##3}\VANthebibliography}

\usepackage{graphicx}	% Including figure files
\usepackage{amsmath}	% Advanced maths commands
\usepackage{gensymb}
\usepackage{hyperref}
\usepackage{ulem}
\usepackage{caption}
\usepackage{footmisc}

\title[A dynamical survey of the trans-Neptunian region II.]{A dynamical survey of the trans-Neptunian region II.: On the nature of chaotic diffusion}

\author[E. K\H ov\'ari et al.]{
E. K\H ov\'ari,$^{1,5}$\thanks{E-mail: kovari.emese@ttk.elte.hu (EK)}
E. Forgács-Dajka,$^{1,4,5}$
T. Kov\'acs,$^{3}$
Cs. Kiss,$^{2,6,7}$
Zs. S\'andor$^{1,6,7}$
\\
$^{1}$ELTE E\"otv\"os Lor\'and University, Institute of Physics and Astronomy, Department of Astronomy, H-1117 Budapest, P\'azm\'any P\'eter s\'et\'any 1/A, Hungary\\
$^{2}$ELTE E\"otv\"os Lor\'and University, Institute of Physics and Astronomy, H-1117 Budapest, P\'azm\'any P\'eter s\'et\'any 1/A, Hungary\\
$^{3}$ELKH--ELTE Extragalactic Astrophysics Research Group, E\"otv\"os Lor\'and University, H-1117 Budapest, P\'azm\'any P\'eter s\'et\'any 1/A, Hungary\\
$^{4}$ELKH--SZTE Stellar Astrophysics Research Group, H-6500 Baja, Szegedi út, Kt. 766, Hungary\\
$^{5}$Wigner Research Centre for Physics, H-1525 Budapest, P.O. Box 49, Hungary\\
$^{6}$Konkoly Observatory, Research Centre for Astronomy and Earth Sciences, H-1121 Budapest, Konkoly Thege Miklós út 15-17, Hungary\\
$^{7}$CSFK, MTA Centre of Excellence, H-1121 Budapest, Konkoly Thege Miklós út 15-17, Hungary
}

\date{Accepted 2023 May 29. Received 2023 May 29; in original form 2023 February 20}

\pubyear{2023}

\begin{document}
\label{firstpage}
\pagerange{\pageref{firstpage}--\pageref{lastpage}}
\maketitle

\begin{abstract}
On long enough timescales, chaotic diffusion has the potential to significantly alter the appearance of a dynamical system. The solar system is no exception: diffusive processes take part in the transportation of small bodies and provide dynamical pathways even for the distant trans-Neptunian objects to reach the inner solar system. In this Letter, we carry out a thorough investigation of the nature of chaotic diffusion. We analyze the temporal evolution of the mean squared displacement of ten thousand ensembles of test particles and quantify in each case the diffusion exponent (enabling the classification between normal, sub-, and super-diffusion), the generalized diffusion coefficient, and a characteristic diffusion timescale, too. This latter quantity is compared with an entropy-based timescale, and the two approaches are studied in light of direct computations as well. Our results are given in the context of two-dimensional maps, thereby facilitating the understanding of the relationship between the typical phase space structures and the properties of chaotic diffusion.
\end{abstract}

\begin{keywords}
celestial mechanics -- chaos -- diffusion -- methods: numerical -- Kuiper belt: general
\end{keywords}

%%%%%%%%%%%%%%%%%%%%%%%%%%%%%%%%%%%%%%%%%%%%%%%%%%

%%%%%%%%%%%%%%%%% BODY OF PAPER %%%%%%%%%%%%%%%%%%

\section{Introduction}
\label{sec:introduction}

The problem of determining the degree of global (in)stability of a planetary system is a frequently visited yet challenging problem in dynamical astronomy. Although dynamical systems are deterministic, they do exhibit seemingly irregular motion \citep{geisel1982}. This chaotic behaviour, however, stems not from the presence of random external forces (as would be the case in a stochastic system) but from non-linearities in the equations of motion. A criterion for chaoticity in dynamical systems is the extreme sensitivity to initial conditions (ICs) which practically implies unpredictability, for the ICs cannot be measured with infinite precision.

For the quantification of chaos, several indicators were introduced in the course of time (see e.g.: mLCE - \cite{benettin1980}; FLI - \cite{Froeschle1997}; MEGNO - \cite{Cincotta2000}; SALI/GALI - \cite{Skokos2001,Skokos2004}; RLI - \cite{Sandor2004}). They suffice for the quick and robust differentiation between phase space regions of fundamentally distinct dynamical characteristics, yet they lack the capacity of providing information about global instability. Alas, despite the great number of works attempting to establish a relationship between chaos and global instability, no general results were given so far \citep[see][for a review]{Shevchenko2020}.

In this regard, chaotic diffusion may be a key quantity, provided that its quantification allows for the prediction of global instability. By chaotic diffusion, we refer to the erratic motion of a phase point in the chaotic regions of the phase space. If considering an ensemble of initially neighbouring such phase points, most often we find that their mean squared displacement (MSD) grows linearly with time. This process is the Brownian motion-like normal diffusion. Deviations from the linear scaling are termed as anomalous diffusion (which can either be slower -- sub-diffusion -- or faster -- super-diffusion -- than the normal diffusion). Although it seems less frequent in nature, anomalous diffusion is not restricted to artificial or exceedingly complex systems but manifests both in simple systems \citep[see, among others,][]{Leonel2009,Livorati2012,Harsoula2018} and e.g. in planetary systems, too \citep[][etc.]{cordeiro2006,marti2016}. Numerical treatment of normal diffusion may be easier, but the above examples underline the importance of broadening our understanding of all three types of diffusion.

In the case of the solar system, chaotic diffusion was first detected by \cite{laskar1994} in the eccentricity of Mercury. Since then, several other works \citep[for a large-scale statistical study see e.g.][]{laskar2008} drew attention to its substantial role in shaping the dynamical structure. Since chaotic diffusion is not limited to single particles but is inherent to all the bodies as some intrinsic dynamical property, on long enough timescales, it can alter the appearance of the entire system. For instance, the depletion of the 2:1 Kirkwood gap in the main asteroid belt is attributed to diffusive forces \citep{morbidelli1996,tsiganis2002}, as is the origin of short-period comets \citep{morbidelli1997}, and of Centaurs and scattered disk objects \citep{tiscareno2009}. The present-day overpopulation of the Plutinos (trans-Neptunian objects (TNOs) in the 3:2 mean-motion resonance (MMR) with Neptune) in comparison with that of the Twotinos (TNOs in the 2:1 MMR with Neptune) can likewise be understood by different rates of chaotic diffusion \citep{tiscareno2009}. In these previous studies, however, the possibly anomalous nature of the chaotic diffusion is not considered. The aim of this paper is to give an overall description of diffusive processes -- normal and anomalous alike. We concentrate on the innermost, $ 34 - 40 $ AU region of the trans-Neptunian space -- that includes three first-order MMRs, the most populated 3:2 MMR among them --, and carry out a comprehensive survey of the chaotic diffusion. We present our results in the framework of two-dimensional heat maps to ease the visualization of the diffusion-wise different phase space domains.

The paper is organized as follows. In Section~\ref{sec:measure_of_the_chaotic_diffusion_by_means_of_MSD_analysis}, first, we introduce the method of the MSD analysis to quantify chaotic diffusion and its parameters, then, after outlining the details of our computations, we provide the related results in Section~\ref{subsec:diffusion_maps}. Thereafter, in the first part of Section~\ref{sec:independent_measures_of_the_chaotic_diffusion}, we present an alternative, entropy-based measure of the chaotic diffusion, which is followed, in the second part, by the results of direct, long-term numerical integrations as independent reference for the indirect computations. Lastly, Section~\ref{sec:summary} is a brief summary of the work.

\section{Measure of the chaotic diffusion by means of MSD analysis}
\label{sec:measure_of_the_chaotic_diffusion_by_means_of_MSD_analysis}

\subsection{The method}
\label{subsec:the_method}

A common approach to quantifying chaotic diffusion and its parameters is through the measurement of the MSD of a random variable $ \mathbf{x} $ (an ensemble of $ n_\mathrm{part} $ trajectories in our case, $ \mathbf{x} \equiv (\mathbf{x}_i)_{i=1}^{n_\mathrm{part}} $), starting from $ \mathbf{x}(0) $ at time $ t = 0 $. The MSD of $ \mathbf{x} $ is by definition: $ \mathrm{MSD}_\mathbf{x}(t) \equiv \left\langle \left(\mathbf{x}(t) - \mathbf{x}(0)\right)^2 \right\rangle $, where $ \langle . \rangle $ denotes an ensemble average. According to the generalized Einstein relation, the MSD is proportional to the $ \alpha $th power of time as
\begin{equation}
	\mathrm{MSD}_\mathbf{x}(t) = 2dD_\alpha t^\alpha,
	\label{eq:MSDdiff}
\end{equation}
where $ d $ is the dimensionality of $ \mathbf{x} $, $ D_\alpha $ is the generalized diffusion coefficient of dimension (phase space) $ \mathrm{distance}^2 \cdot \mathrm{time}^{-\alpha} $, and the diffusion exponent $ \alpha $ determines whether the process is of normal diffusion ($\alpha = 1 $), or if it is categorized as sub-diffusive ($ 0\leq \alpha < 1 $ -- slow diffusion) or as super-diffusive ($ 1 < \alpha $ -- fast diffusion). (Usually, the ballistic limit $ \alpha = 2 $ is considered as the upper limit.)

In the case of normal diffusion, the only unknown in Equation~\eqref{eq:MSDdiff} is the diffusion coefficient $ D_\alpha =:D_1 $, its numerical derivation is not complicated. For anomalous diffusion, both $ \alpha $ and $ D_\alpha $ are to be determined, thus another approach is needed. A simple concept is to take the logarithm of both sides of Equation~\eqref{eq:MSDdiff} and to fit the $ \log_{10} (t) \mapsto \log_{10} (\mathrm{MSD}_\mathbf{x}(t)) $ linear function \citep[a similar approach can be found in the papers of][]{cordeiro2005,cordeiro2006}. The slope of the fit then yields the diffusion exponent $ \alpha $, whereas the vertical intercept gives $ \log_{10}(2dD_\alpha) $, from where the coefficient $ D_\alpha $ is easily obtainable.

Once $ \alpha $ and $ D_\alpha $ are known, a characteristic diffusion time
\begin{equation}
    \tau := \left(\frac{\Delta^2}{2dD_\alpha}\right)^{\left(1/\alpha\right)}
    \label{eq:tau}
\end{equation}
can be defined, by rearranging Equation~\eqref{eq:MSDdiff}. $ \Delta^2 $ in the expression is a specific mean squared displacement of Equation~\eqref{eq:MSDdiff}, which is related to some stability criterion of free choice. Our selection for $ \Delta^2 $ is specified in Section~\ref{subsec:free_parameters_of_the_MSD_analysis}.

\subsection{Numerical integrations}
\label{subsec:numerical_integrations}

For the detailed investigation of how the chaotic diffusion manifests in the phase space of the outer solar system, we integrated the equations of motion of two hundred thousand test particles and then constructed two-dimensional heat maps depicting the parameters of the diffusion.

The numerical integration of the test particles \citep[for the full description see][]{forgacsdajka2023} was carried out in a barycentric coordinate system containing the four giant planets, Pluto, and the Sun. The total integration time was set to $ 2\cdot10^5 $~yrs (corresponding approximately to 1200 Neptune periods) and the sampling time step to $ 100 $~yrs. The mass-less test particles were placed in a rectangular grid of the semi-major axis ($ a $) - eccentricity ($ e $) parameter plane, and the initial orbital elements of the major planets and Pluto, as well as those of the Sun, were downloaded from the JPL Horizons database\footnote{https://ssd.jpl.nasa.gov/horizons/} at the epoch 01 January 2022, 0h UT.

As for the choice of the $ a $-range of the grid, we considered the interval $ 34 \text{ AU} \leq a \leq 40 \text{ AU} $, which incorporates three of the most important first-order MMRs (the 5:4 resonance close to the orbit of Neptune at 30 AU, the distant 3:2 resonance of the Plutinos, and the 4:3 resonance in between). The eccentricities were selected from the interval $ 0 \leq e \leq 0.6 $. This two-dimensional section $ [34 \ \mathrm{AU}, 40 \ \mathrm{AU}] \times [0, 0.6] $ of the phase space was then divided into $ 100\times100 $ (``big'') grid cells. Within each cell, there is an ensemble of 20 ICs, distributed again in a rectangular manner (that is, the ``big'' cells were divided into $ 5\times4 $ ``small'' cells -- in $ a $ and in $ e $, respectively). Altogether, such a construction sets the initial actions of $ 2\cdot10^5 $ test particles. The angle-like orbital elements were then chosen randomly from the ranges $ 0\degree \leq I \leq 30\degree $ (inclination), $ 0\degree \leq \omega \leq 360\degree $ (argument of perihelion), $ 0\degree \leq M \leq 360\degree $ (mean anomaly) -- assuming uniform distributions, and the longitudes of ascending node we set to $ \Omega = 0\degree $.

\subsection{Free parameters of the MSD analysis}
\label{subsec:free_parameters_of_the_MSD_analysis}

The numerical integrations having been completed, we carried out the MSD analysis (detailed in Section~\ref{subsec:the_method}) for the ten thousand ensembles of test particles. In the following, we specify our selection for the free parameters of the method.

As for the choice of the phase space variable $ \mathbf{x} $ in Equation~\eqref{eq:MSDdiff}, it proved to be advantageous to switch from the classical orbital elements to the Delaunay variables, for they remain well-defined and non-singular for very nearly circular ($ e \approx 0 $) and/or flat ($ I \approx 0\degree $) orbits. We used the Delaunay actions (of dimension $ \mathrm{AU}^2/\mathrm{yr} $)
\begin{equation}
    L = \sqrt{\mu a}, \qquad G = L\sqrt{ ( 1-e^2 )}, \qquad H = G\cos(I)
\end{equation}
($ \mu $ is the standard gravitational parameter of the Sun in astronomical units) to construct the three-dimensional ($ d = 3 $) $ \mathbf{x} = (L, G, H) $.

The linear fit -- with least squares approximation -- of the $ \left(\log_{10}(\mathrm{MSD}_\mathbf{x}(t)), \log_{10}(t) \right) $ curves yields the fitting parameters $ \alpha $ (diffusion exponent) and $ D_\alpha $ (diffusion coefficient). The goodness of fit was checked by computing the coefficients of determination
\begin{equation}
    R^2 = 1 - \frac{S\!S_\mathrm{res}}{S\!S_\mathrm{tot}},
    \label{eq:R2}
\end{equation}
where $ S\!S_\mathrm{res} $ stands for the residual sum of squares (the sum of the squared differences between the data points and the fitted values) and $ S\!S_\mathrm{tot} $ for the total sum of squares (the sum of the squared differences between the data points and their mean value).

Lastly, we needed to choose a physically plausible $ \Delta^2 $ in order to compute the diffusion timescale $ \tau $ in Equation~\eqref{eq:tau}. We linked it to orbital stability through Hill's criterion \citep{gladman1993} $ C_\mathrm{H} = 2\sqrt{3} R_\mathrm{H} $, where $ R_\mathrm{H} $ is the Hill radius of Neptune. As long as the minimal separation of a major planet (meaning now Neptune) and a minor one (a test particle) is larger than $ C_\mathrm{H} $, the stability of the small body is ensured. We took $ \Delta a := C_\mathrm{H} $ to construct (through differentiating with respect to $ a $)
\begin{equation}
    \label{eq:deltas}
    \Delta L = \frac{\Delta a}{2}\sqrt{\frac{\mu}{a}},
    \quad \ \ \ \
    \Delta G = \Delta L \sqrt{(1-e^2)}, 
    \quad \ \ \ \
    \Delta H = \Delta G \cos(I).
\end{equation}
(We equated the nominal values of $ a $, $ e $, and $ I $ in Equations~\eqref{eq:deltas} with their initial ones -- which refer to the mean values in the case of ensembles.) With the above stability criteria for the Delaunay actions, we set
\begin{equation}
    \Delta^2 := (\Delta L)^2 + (\Delta G)^2 + (\Delta H)^2,
    \label{eq:Delta_sq}
\end{equation}
and hence the linkage of $ \tau $ to Hill's stability.

\subsection{Diffusion maps}
\label{subsec:diffusion_maps}

In the present subsection, we provide two-dimensional heat maps of the $ 34-40 $~AU region of the trans-Neptunian space that portray the diffusion exponents $ \alpha $, the diffusion coefficients $ D_\alpha $, and the diffusion timescales $ \tau $. We also enclose the linear fits' coefficients of determination $ R^2 $.

\begin{figure}
    \centering
    \includegraphics[width=\columnwidth]{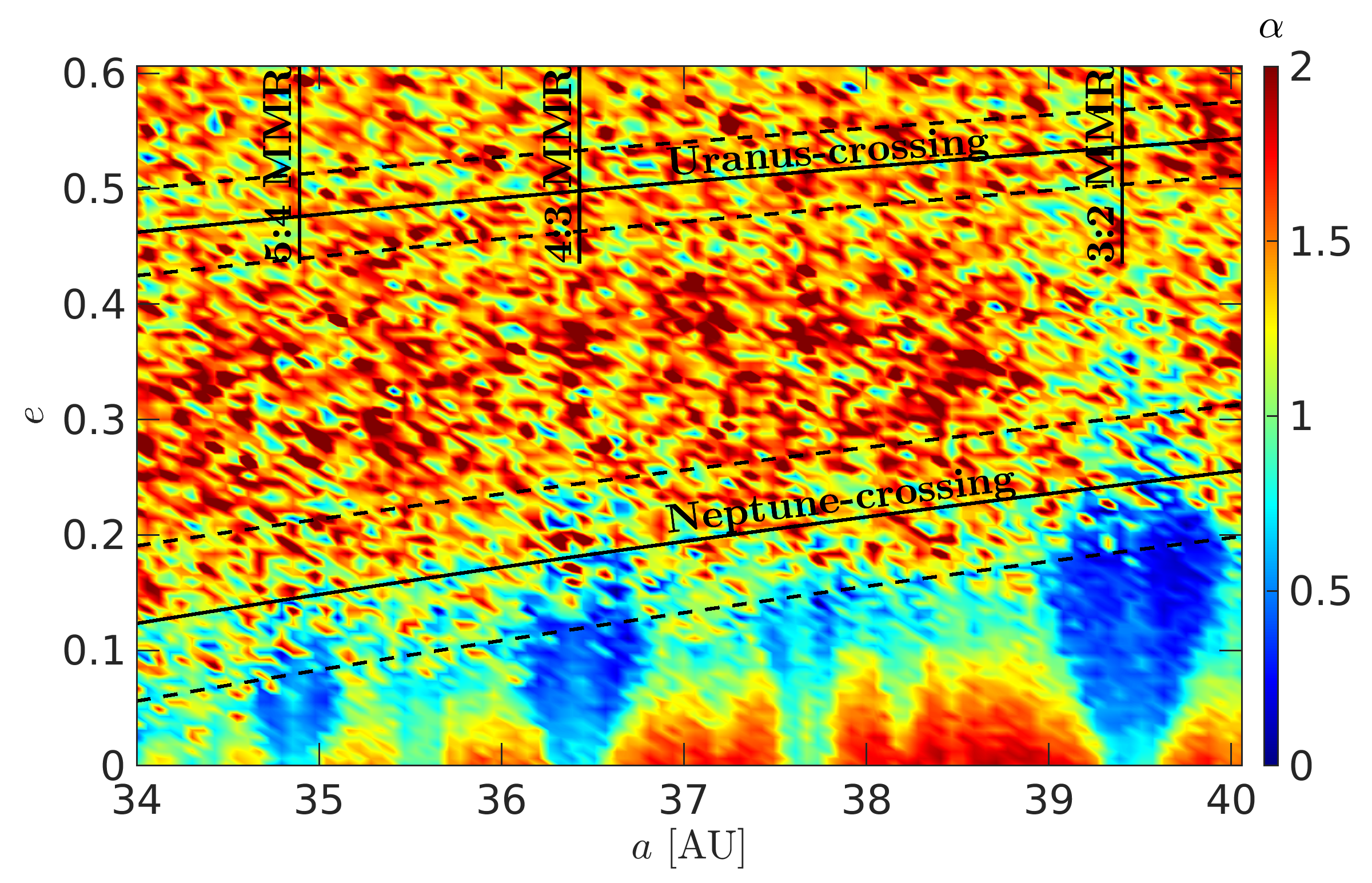}
    \caption{Map of the diffusion exponents $ \alpha $ in Equation~\eqref{eq:MSDdiff} computed from the MSD of $ \mathbf{x} = (L, G, H) $ in the $ 34-40 $ AU region of the trans-Neptunian space (between the eccentricities $ 0 \leq e \leq 0.6 $). Solid lines: $(a,e)$ pairs that result in Neptune- or Uranus-crossing orbits. Dashed lines: three Hill radii distances from the two giant planets. Vertical lines on the top of the figure: location of MMRs.}
    \label{fig:alpha}
\end{figure}

Figure~\ref{fig:alpha} shows the $ \alpha $-map. The colour code is such that from dark blue to dark red, the value of $ \alpha $ increases, that is, bluish shades are assigned to the case of sub-diffusion, the cyan and green hues belong to normal diffusion, and from yellow, we are faced with super-diffusion. (Exponents greater than $ \alpha = 2 $ got the last hue of the colour bar.) The first observation of the figure is that the characteristic V-shapes of the strongest (first-order) MMRs dominate the lower part of the plot. They penetrate to some extent into the upper domains, too, cutting through the region of Neptune-crossing $ (a, e) $ pairs, indicated by the lower solid line with the accompanying dashed lines of three Hill radii distances from the giant planet. In the interior of the resonances, the dynamics is sub-diffusive, with exponents $ \alpha \lesssim 0.5 $. At their boundaries -- where one expects the phenomenon of the so-called stickiness to manifest --, the greenish colours imply normal diffusion ($ \alpha \sim 1 $). The super-diffusion with $ \alpha \gtrsim 1.2 $ appears above the Neptune- (and Uranus-) crossing regions at higher eccentricities, but in the regime of the nearly circular orbits, too, where its presence is somewhat surprising. As an explanation, we proceed to discuss the role of the other fitting parameter $ D_\alpha $.

\begin{figure}
    \centering
    \includegraphics[width=\columnwidth]{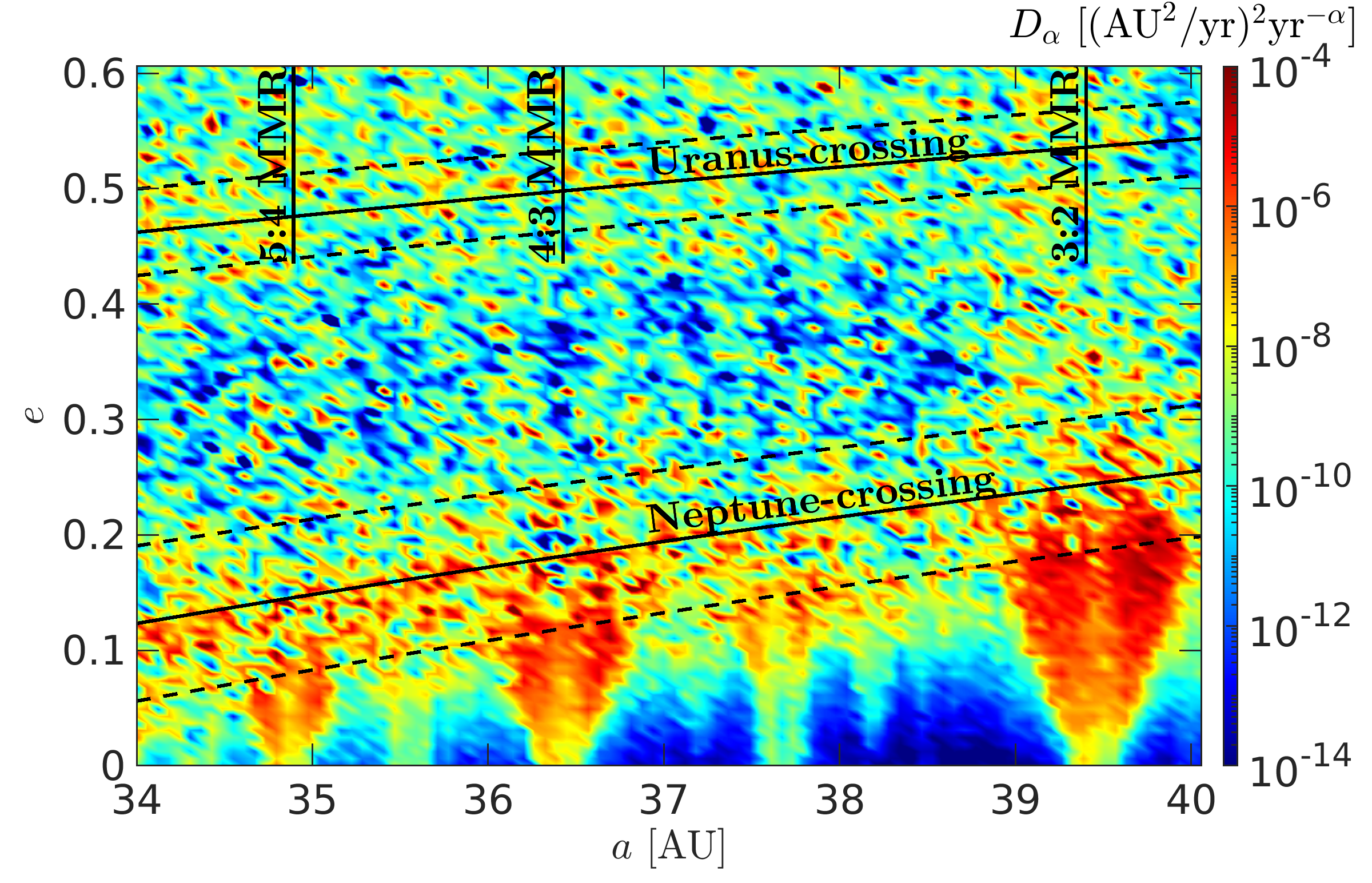}
    \caption{The same as Figure~\ref{fig:alpha}, but depicting the diffusion coefficients $ D_\alpha $ in Equation~\eqref{eq:MSDdiff}.}
    \label{fig:D}
\end{figure}

The map of $ \alpha $ in itself should not be viewed as a complete description of the chaotic diffusion, for the diffusion coefficient $ D_\alpha $ can fundamentally alter the final picture. Figure~\ref{fig:D} represents the map of $ D_\alpha $. Again, the colour scale ranges from dark blue to dark red, appointed now to the interval $ 10^{-14} \leq D_\alpha \leq 10^{-4} \ (\mathrm{AU}^2/\mathrm{yr})^2 \mathrm{yr}^{-\alpha} $. (Values outside this range got the last hues of the colour bar.) The main structures of the previous figure reveal here, too, but with ``inverse'' properties. The lowest values of the diffusion coefficient show up in the circular orbits' region -- exactly where large $ \alpha $ exponents appeared in Figure~\ref{fig:alpha}. As for the MMRs, their associated $ D_\alpha $ parameters are the maximal ($ \sim 10^{-4} - 10^{-6} \ (\mathrm{AU}^2/\mathrm{yr})^2 \mathrm{yr}^{-\alpha} $). In the rest of the present phase space section, intermediate values are read. Therefore, the combination of Figures~\ref{fig:alpha} and \ref{fig:D}, gives instances of both low $ \alpha $ -- high $ D_\alpha $ and high $ \alpha $ -- low $ D_\alpha $ pairs. These two, seemingly opposite, cases can be indicators of similar dynamical outcomes.

\begin{figure}
    \centering
    \includegraphics[width=\columnwidth]{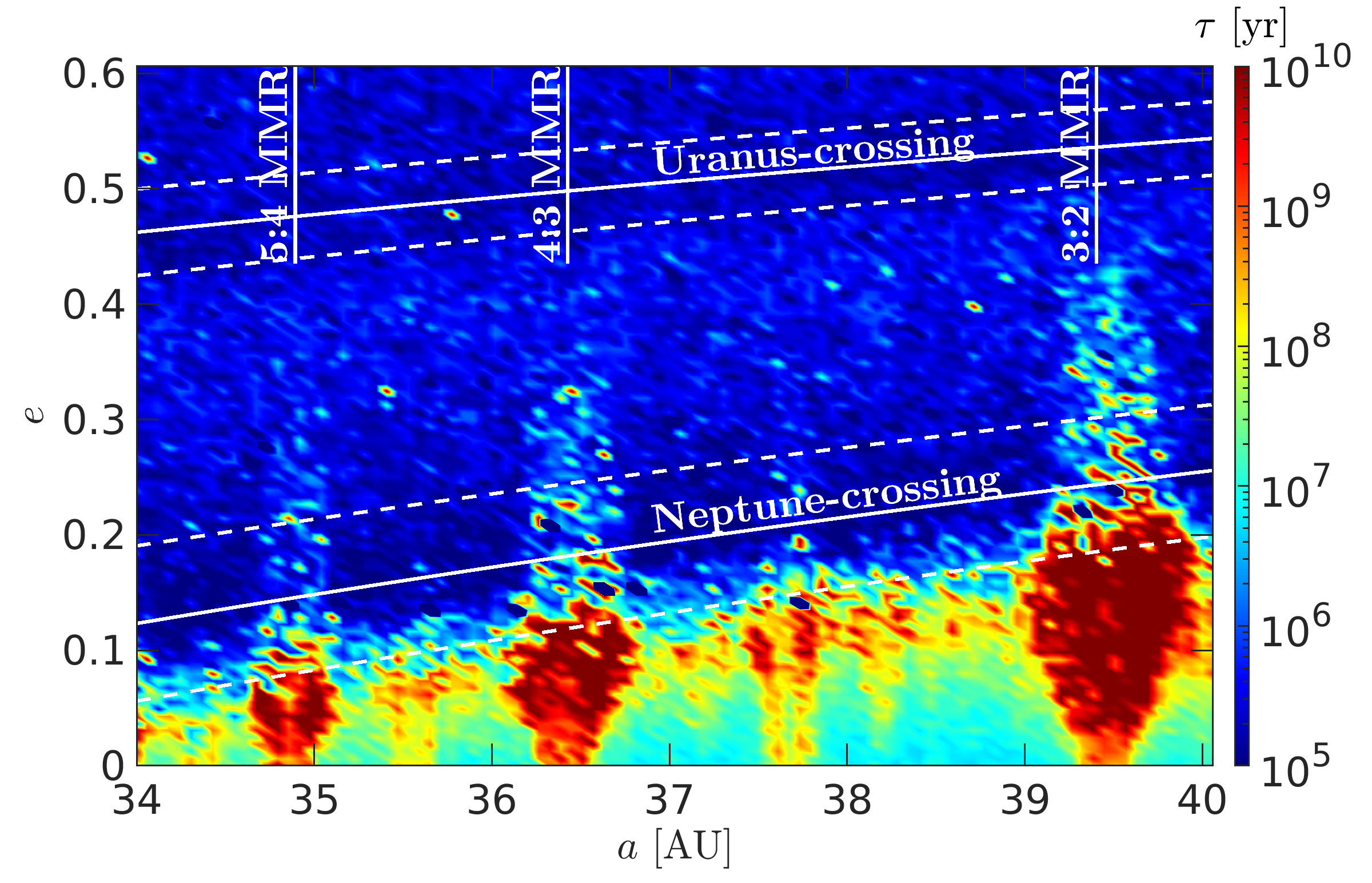}
    \caption{The same as Figure~\ref{fig:alpha}, but depicting the characteristic diffusion timescales $ \tau $ in Equation~\eqref{eq:tau}.}
    \label{fig:tau}
\end{figure}

To further examine the joint role of $ \alpha $ and $ D_\alpha $, we enclose in Figure~\ref{fig:tau} the map of the diffusion times $ \tau $, computed according to Equations~\eqref{eq:tau} and \eqref{eq:Delta_sq} and truncated at the lower and upper thresholds $ 10^5 $ and $ 10^{10} $~yrs, respectively. Although it is important to bear in mind that the predictions for $ \tau $ are based on the finite integration times of $ 2\cdot10^5 $~yrs as well as on the assumption that the diffusive nature remains unaltered later on, the phase space structure is now mostly as expected. In the interior of the MMRs, the diffusion times rise high; the values -- equal to or larger than the age of the Universe -- are nominal but indicate long-term stability nevertheless. In the low-eccentricity regime, the $ \tau $s are slightly smaller, though they still exceed $ 10^7 $~yrs. At and beyond the Neptune-crossing line, however, there is a loss of one or two more orders of magnitude. The lowest values of the figure ($ \tau \sim 10^5 $~yrs) belong to the doubtlessly unstable domains in the vicinity of the giant planets (see the lower and upper pairs of dashed lines), whereas $ \tau $ noticeably increases to $ \sim 10^6 $~yrs in between them.

\begin{figure}
    \centering
    \includegraphics[width=\columnwidth]{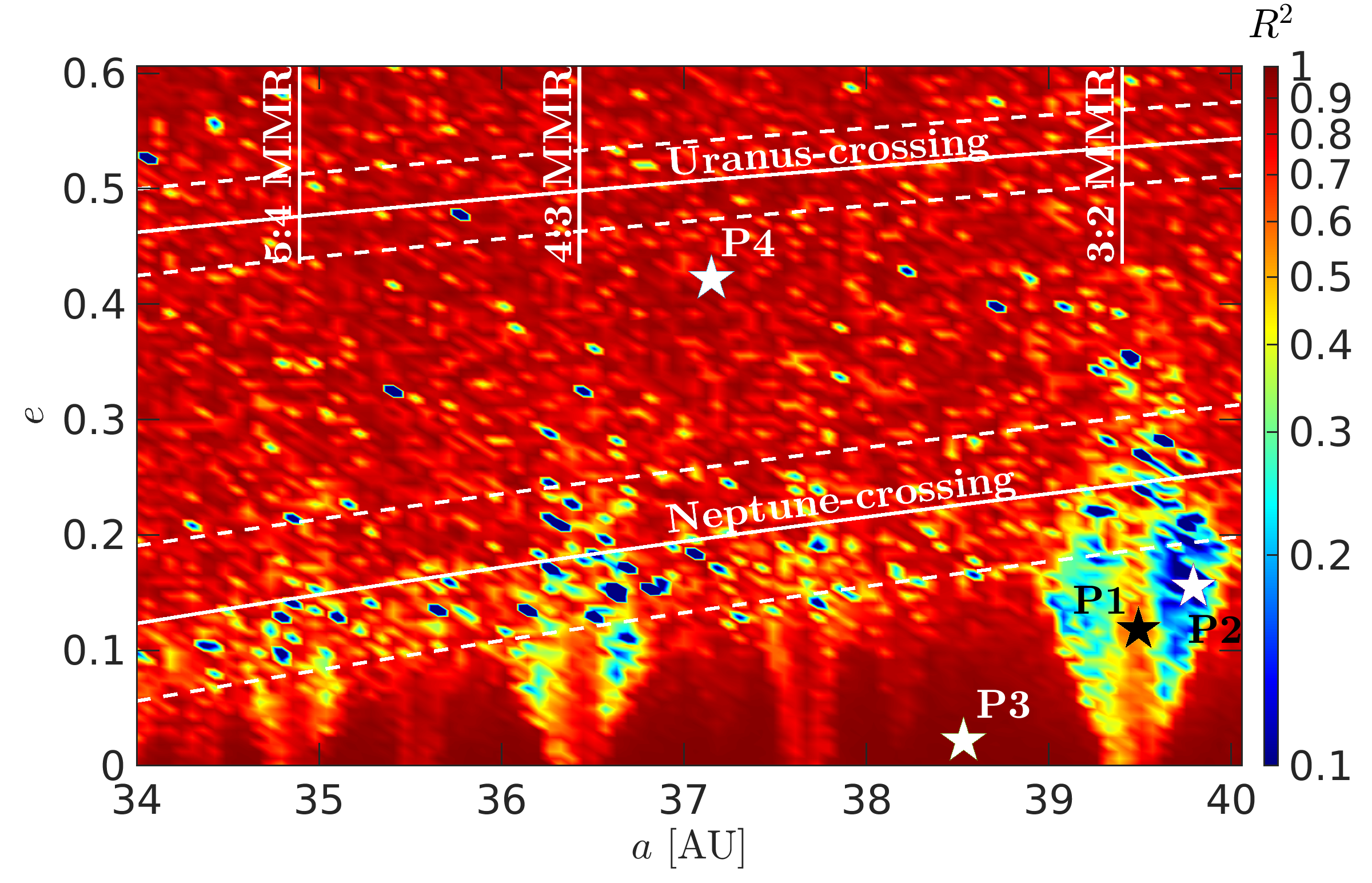}
    \caption{The same as Figure~\ref{fig:alpha}, but depicting the coefficients of determination $ R^2 $ in Equation~\eqref{eq:R2}. 
    The points P1, P2, P3, and P4 -- marked with stars -- are sampling locations for Figure~\ref{fig:fits}.}
    \label{fig:R2}
\end{figure}

\begin{figure}
    \centering
    \includegraphics[width=\columnwidth]{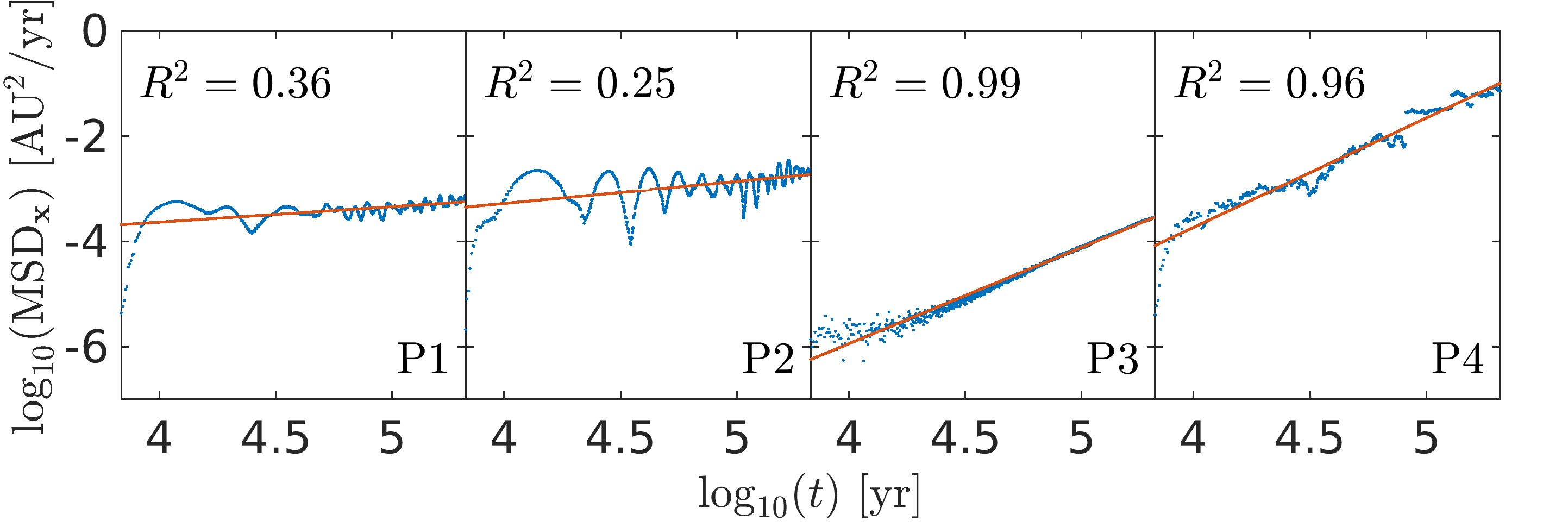}
    \caption{Linear fit (red) of the logarithm of the MSD curves (blue) of the ensembles P1 (first panel), P2 (second panel), P3 (third panel), and P4 (fourth panel) marked with stars in Figure~\ref{fig:R2}. The $ R^2 $ value of the fit is indicated at the top of each panel.}
    \label{fig:fits}
\end{figure}

Let us now remark that the linear fit of the log-log plot of Equation~\eqref{eq:MSDdiff} is, of course, laden with error. For the quantification of the goodness of fit, we subjoin the map of the coefficients of determination $ R^2 $ (see Equation~\eqref{eq:R2}) in Figure~\ref{fig:R2}. The coefficients are generally high ($ \sim 0.8-1 $) in the majority of the figure, except for the interior of the first-order MMRs -- toward the separatrices --, where one finds $ R^2 \lesssim 0.4 $. The possible sources of the error could, on the one hand, be large fluctuations in the MSD curve around a mean value, but changes in the type of diffusion as well. We inspected several individual fits at various locations of the map to settle the question and found the former error source to be relevant. Four representative fits are shown in Figure~\ref{fig:fits} and the locations of the corresponding ensembles are marked with stars in Figure~\ref{fig:R2} (see the points P1, P2, P3, P4). At the exact centre of the MMRs (first panel of Figure~\ref{fig:fits}, P1), the libration amplitude of the resonant argument is very small and so is the change in the orbital elements. But as we move toward the separatrices (second panel of Figure~\ref{fig:fits}, P2), it grows gradually. Accordingly, the fluctuations of the MSD curves increase, too, which impairs the goodness of fit. Note though that the trend of the line is not affected and it clearly indicates sub-diffusion in both cases. (For comparison, we added two more examples (third and fourth panels of Figure~\ref{fig:fits}, P3 and P4), selected from the low- and high-eccentricity regimes, respectively.) An important message of the above observations is that even though the MMRs themselves are responsible for the large (periodic) changes in the orbital elements of the particles, they also offer them phase protection, thereby preventing close encounters with the major planets.

The above results are based on Equation~\eqref{eq:MSDdiff}. It is noteworthy though that, strictly, the formula is valid only for (nearly) isotropic diffusion. We performed all the above computations separately for $ L $, $ G $, and $ H $ and found that in the majority of the $ [34 \ \mathrm{AU}, 40 \ \mathrm{AU}]\times[0, 0.6] $ section, isotropy holds. The only exception is the low-eccentricity, non-resonant region, where the overall diffusion is dominated by that in $ H $ ($ \mathrm{MSD}_{(L, G, H)} \approx \mathrm{MSD}_H $). However, it also means that if in this domain, we were to exclude $ L $ and $ G $ from the computations and keep only $ H $, the results would remain mainly unaltered. These findings underline the important role of the (initial) inclinations (contained by $ H $), an observation already seen in our previous paper \citep{forgacsdajka2023}.

\section{Independent measures of the chaotic diffusion}
\label{sec:independent_measures_of_the_chaotic_diffusion}

In the previous section, we presented a comprehensive survey of the chaotic diffusion in the inner part of the trans-Neptunian space with recourse to the MSD analysis. This section is devoted to a comparison of the results with an independent, entropy-based measure, followed by the presentation of direct, long-term computations.

\subsection{A Shannon-entropy-based method}
\label{subsec:a_shannon_entropy_based_method}

The Shannon entropy, introduced originally in information theory \citep{shannon1949}, has widely been proven \citep[see e.g.][]{cincotta2012,cincotta2018,giordano2018,beauge2019,cincotta2021march,kovari2022} to be efficiently applicable in dynamical astronomy, too. The derivative of the entropy gives an account of the evolution of the ``spread'' of a (/an ensemble of) phase space trajectory (/trajectories), and thus the rate of chaotic diffusion can be estimated.

For the theoretical background of the Shannon entropy formalism when applied in dynamical astronomy, we refer to the above-cited works and provide here only the technical aspects of our computations. To facilitate comparison with our previous figures, we used the same $ [34 \ \mathrm{AU}, 40 \ \mathrm{AU}] \times [0, 0.6] $ (initial values of $ a $ and $ e $) phase space region as before and worked with the same set of ICs: $ 100\times100 $ ensembles of 20 test particles in each (see Section~\ref{subsec:numerical_integrations}). Then we created secondary partitions around each ensemble to track the ``diffusion'' of the trajectories as time passes. We note here that while in earlier works two-dimensional partitions were used (mostly in the $ (a, e) $ or $ (L, G) $ planes), we extend the computations to a third dimension to better suit the needs of the high complexity of the outer solar system. Therefore, the properties of our secondary partitions are as follows: dimension: $ d = 3 $; selected variables: $ L, G, H $; centre: computed from the central $ a $, the central $ e $, and $ \langle I \rangle = 15\degree $; size: $ [-\Delta L, \Delta L]\times[-\Delta G, \Delta G]\times[-\Delta H, \Delta H] $ (see Equations~\eqref{eq:deltas}); number of cells: $ r := r_L \cdot r_G \cdot r_H = 100^3 $. With such initialization, the Shannon entropy of a given ensemble reads
\begin{equation}
    S = - \sum_{k = 1}^r p_k \ln(p_k),
    \label{eq:shannon}
\end{equation}
where $ p_k := n_k / \bar{N} $ is the probability that at a given time step the subsequent trajectory points of the ensemble will fall into the $ k $th cell of the partition ($ \bar{N} = N \cdot n_\mathrm{part} $, $ N = 2000 $ is the number of integrational points (i.e. trajectory points) of a given member of the ensemble, $ n_\mathrm{part} = 20 $ is that of the trajectories in one ensemble, and $ n_k $ is that of the trajectory points in the $ k $th cell of the partition). 

From the temporal evolution of the (normalized) Shannon entropy (i.e. $ S \equiv S/\ln(r) $ hereafter), the (global) diffusion coefficient $ D_S $ is obtained as
\begin{equation}
    D_S = \left\langle\frac{(A_{\mathrm{max}} - A_{\mathrm{min}})^2}{r} r_0 \frac{\mathrm{d}S}{\mathrm{d}t}\right\rangle,
    \label{eq:DS}
\end{equation}
where $ A_\mathrm{max} $ and $ A_\mathrm{min} $ are the extrema of the Euclidean norm $ A \equiv \sqrt{L^2 + G^2 + H^2} $, $ r_0 $ gives the number of non-empty cells, and $ \langle . \rangle $ denotes averaging over discrete time intervals. $ \mathrm{d}S / \mathrm{d}t $ was approximated numerically by using the backward finite difference schema.

From $ D_S $, an alternative diffusion time
\begin{equation}
	\tau_S = \frac{\Delta^2}{2dD_S}
	\label{eq:tauS}
\end{equation}
can be formulated (with the same $ \Delta^2 $ as in Equation~\eqref{eq:Delta_sq}).

The construction of $ D_S $ (and $ \tau_S $) by e.g. \cite{cincotta2021febr} is such that the entropy technique is applicable to anomalous diffusion, too. As for the potential anisotropy, the authors introduce in the formula of $ \tau_S $ a numerical multiplier $ K $ that takes into account the diffusion path in the phase space. Yet since this constant is of the order of unity, it does not affect the results notably.

\begin{figure}
    \centering
    \includegraphics[width=\columnwidth]{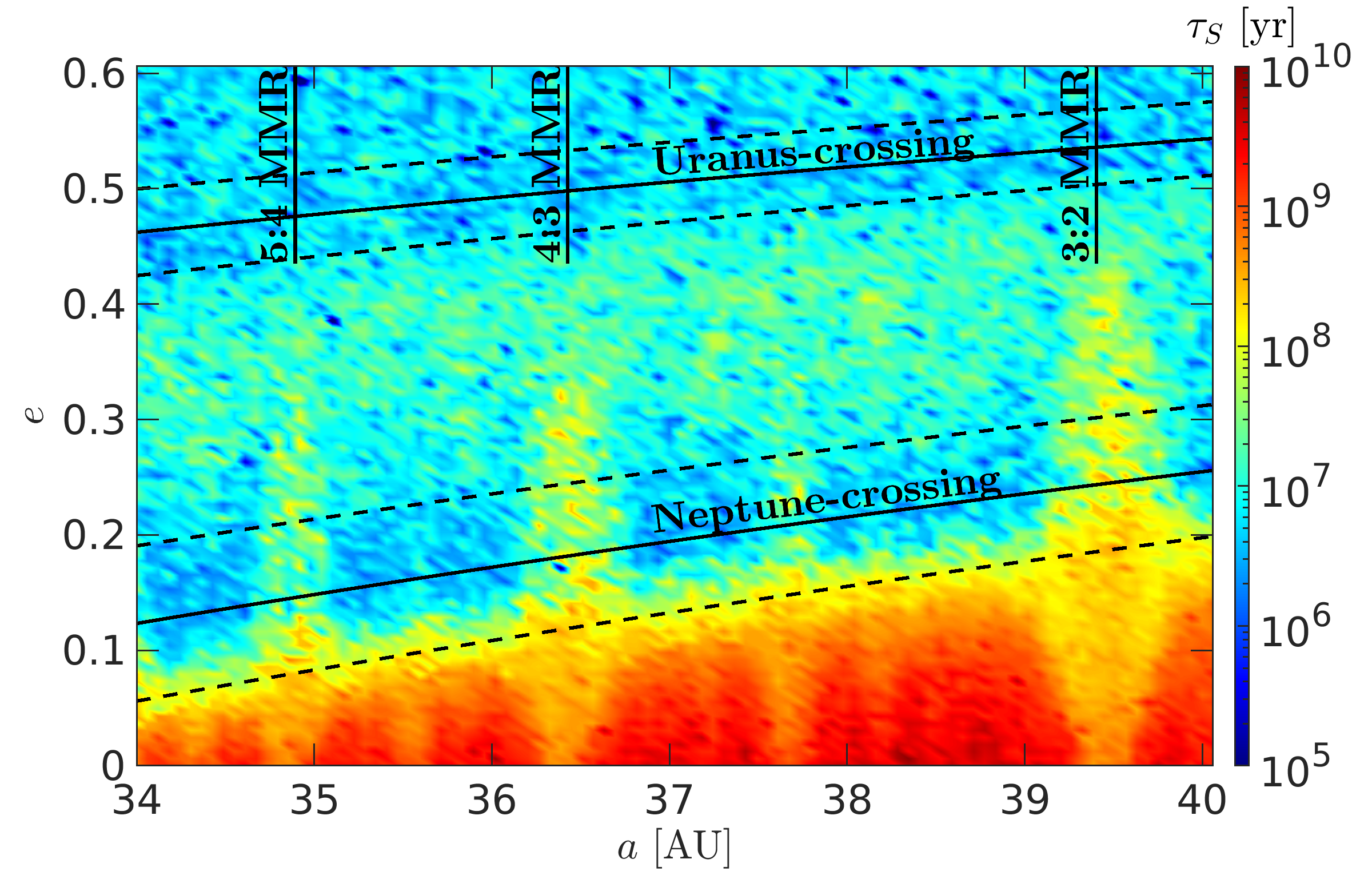}
    \caption{The same as Figure~\ref{fig:alpha}, but depicting the characteristic timescales $ \tau_S $ in Equation~\eqref{eq:tauS}.}
    \label{fig:tauS}
\end{figure}

The map of $ \tau_S $ is displayed in Figure~\ref{fig:tauS}. Both the structure and the colouring of the figure are identical to those of Figure~\ref{fig:tau} hence facilitating the comparison. The main phase space patterns, seen previously, develop here, too; however, with certain deviations of $ \tau_S $ from $ \tau $. While the lowest values of $ \tau_S \sim 10^6-10^7 $~yrs are observed again near and in between the Neptune- and Uranus-crossing regions, we find that the previous results are somewhat overestimated here. As for the lower parts of the figure, the characteristic times of the non-resonant and resonant domains seem to appear in an inverse manner as compared to Figure~\ref{fig:tau}: $ \tau > \tau_S $ inside the MMRs, and $ \tau < \tau_S $ outside of them.

We see, therefore, that the two timescales $ \tau $ and $ \tau_S $ yield non-negligible differences. To decide which one prints a more accurate picture of the dynamical reality of the outer solar system, we present in the following subsection the results of long-term numerical integrations and derive direct estimates of the diffusion.

\begin{figure}
    \centering
    \includegraphics[width=\columnwidth]{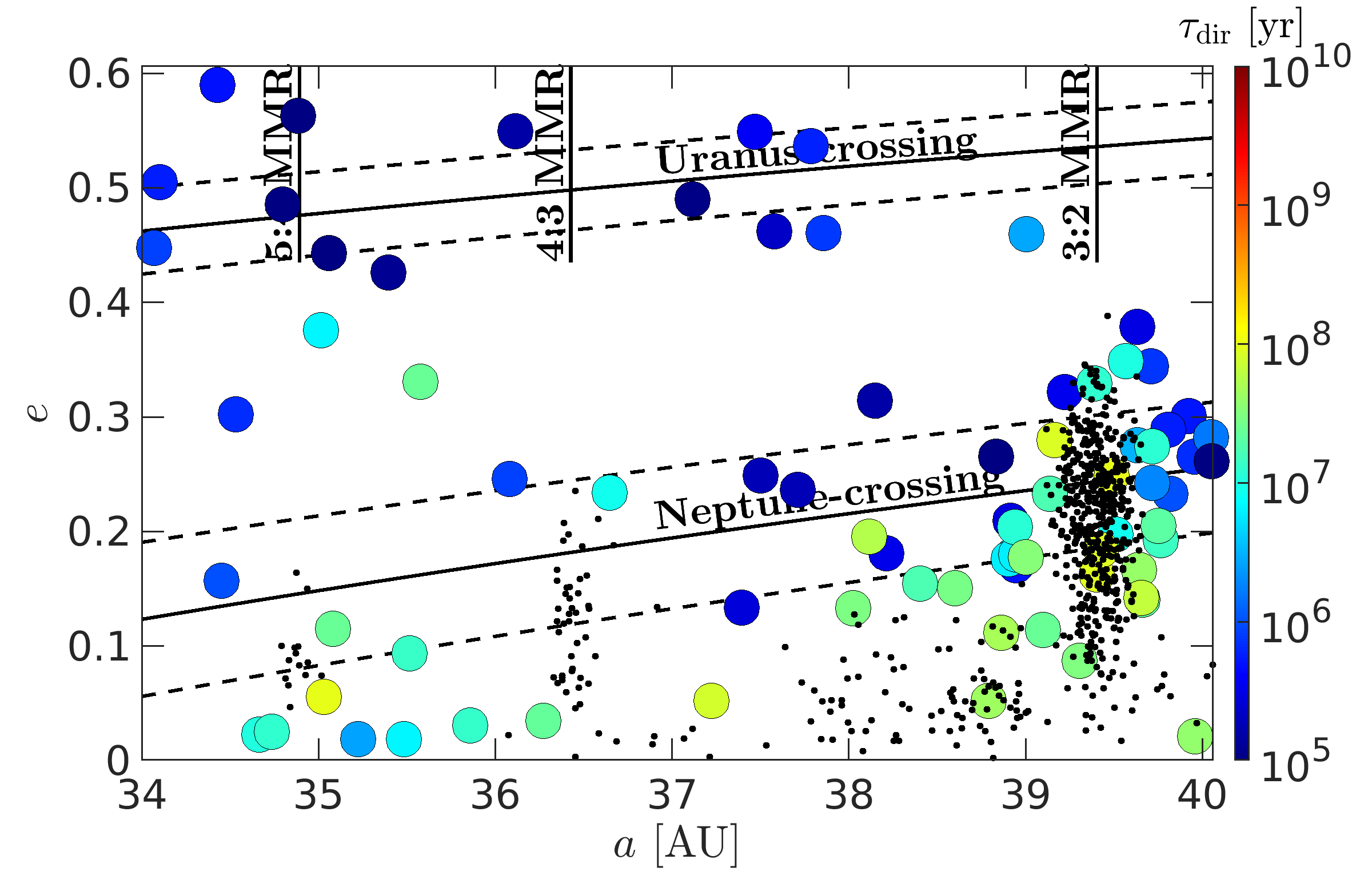}
    \caption{Direct diffusion timescales $ \tau_\mathrm{dir} $ of real TNOs with initial $ (a, e) $ pairs in the range $ [34 \ \mathrm{AU}, 40 \ \mathrm{AU}] \times [0, 0.6] $. Black dots: TNOs with $ \tau_\mathrm{dir} > 10^8 $~yrs. The notations and the colour scale of the figure are the same as those of Figures~\ref{fig:tau} and \ref{fig:tauS}.}
    \label{fig:tau_direct}
\end{figure}

\subsection{Direct results for comparison}
\label{subsec:direct_results_for_comparison}

In our previous work \citep{forgacsdajka2023}, we carried out $ 10^8 $~yr-long numerical integrations of real, though mass-less, TNOs. Here we take advantage of this data set to provide direct estimates of the characteristic timescales of diffusion.

We construct a ``direct diffusion timescale'' $ \tau_\mathrm{dir} $ to capture the time when the Euclidean norm of a given TNO's initial position $ (L, G, H) $ is increased by $ \Delta $ (the square root of $ \Delta^2 $ in Equation~\eqref{eq:Delta_sq}). In Figure~\ref{fig:tau_direct}, at the initial $ (a, e) $ values of those 737 TNOs that fall in the range $ [34 \ \mathrm{AU}, 40 \ \mathrm{AU}] \times [0, 0.6] $, the colours indicate the $ \tau_\mathrm{dir} $ times. TNOs with $ \tau_\mathrm{dir} > 10^8 $~yrs are marked with black dots, whereas the rest are coloured according to the colour coding of Figures~\ref{fig:tau} and \ref{fig:tauS}.

The 5:4, 4:3, and 3:2 first-order MMRs host TNOs of the former category, i.e., with $ \tau_\mathrm{dir} > 10^8 $~yrs. The large values of $ \tau $ in these regions of Figure~\ref{fig:tau} are in good agreement with these results. As for $ \tau_S $, the MMR-values fall exactly in the $ 10^8 $~yrs order of magnitude range, but they do not or only slightly exceed it.

In the non-resonant, high-eccentricity zones, $ \tau_\mathrm{dir} \sim 10^5-10^6 $~yrs, which is closer to the observed values of $ \tau $ than to those of $ \tau_S \sim 10^7 $~yrs.

In the non-resonant, low-eccentricity domain, we see some tens of TNOs with $ \tau_\mathrm{dir} \sim 10^7-10^8 $~yrs, although even more with $ \tau_\mathrm{dir} > 10^8 $~yrs. Here, it is difficult, therefore, to ``announce a winner'', for both $ \tau $ and $ \tau_S $ seem to lack some important component of the reality. We note that this region in question is exactly the same where we found anisotropy in $ L, G, H $ (see Section~\ref{subsec:diffusion_maps}). The results could, therefore, be improved by the adoption of some numerical constants (both in the case of the MSD- and entropy-based approaches), though of different sorts. Yet, the problem deserves deeper investigation, which is out of the scope of this paper.

\section{Summary}
\label{sec:summary}

The chaotic diffusion may seem to be a minor process in comparison with other dynamical effects because of the lengthy timescales it operates on. But over millions or billions of years, its far-reaching implications do become apparent. Acting as a dynamical conveyor belt, it transports bodies throughout the solar system, most notably, escapees of resonant trans-Neptunian objects to the scattered disk population, then to the Centaurs and Jupiter-family comets, to eventually become potentially dangerous asteroids in the inner solar system.

For a better understanding of the general nature of chaotic diffusion -- primarily, whether it is normal or anomalous --, we set out to map its characterizing properties in the innermost part of the trans-Neptunian region by means of MSD analysis. Figures~\ref{fig:alpha}-\ref{fig:tau} are the heat maps of the diffusion exponents, the generalized diffusion coefficients, and characteristic diffusion timescales in the $ 34 - 40 $~AU semi-major axis and $ 0 - 0.6 $ eccentricity ranges. Our results revealed significant deviations from normal diffusion. Weak sub-diffusion dominates the interior of the (5:4, 4:3, and 3:2) MMRs: the test particles here diffuse negligibly and remain under the dynamical protection of the resonances on timescales comparable to the age of the Universe \citep[in accordance with the results of][]{morbidelli1997,tiscareno2009}. As we leave the resonant islands through their separatrices, transitions from sub- to normal, thereafter from normal to super-diffusion take place.

Another approach to measuring chaotic diffusion is through an entropy-based model. The presented method works well to quantify the characteristic timescales of diffusion; however, it does not allow the differentiation between sub-, normal, and super-diffusion. The diffusion timescales resulting from the two methods were, nevertheless, compared as well as studied in view of direct results, the latter based on long-term numerical integration of real TNOs. In the interior of the MMRs, though the two indirect timescales differ, both produce sufficiently long diffusion times as far as the direct results are concerned. In the non-resonant, high-eccentricity domain, the MSD-based approach seems a better estimate, whereas, in the non-resonant, low-eccentricity region, both indirect methods turned out to be in need of improvement.

All in all, this Letter gives a brief overview of the diverse nature of chaotic diffusion in the innermost part of the trans-Neptunian space along with the demonstration of simple yet powerful methods to quantify its parameters. In a forthcoming paper, we intend to extend our studies to the more distant realms of the solar system as well as to further refine the presented techniques.

\section*{Acknowledgements}

We acknowledge the computational resources of the GPU Laboratory of the Wigner Research Centre for Physics. Furthermore, E. K. acknowledges the support of the ÚNKP-22-3 New National Excellence Program of the Ministry for Culture and Innovation from the source of the National Research, Development, and Innovation Fund. C. K. has been supported by the National Research, Development, and Innovation Office (NKFIH), Hungary, through the grant K-138962.

%%%%%%%%%%%%%%%%%%%%%%%%%%%%%%%%%%%%%%%%%%%%%%%%%%
\section*{Data Availability}

The data underlying this article will be shared on reasonable request
to the corresponding author.

%The inclusion of a Data Availability Statement is a requirement for articles published in MNRAS. Data Availability Statements provide a standardised format for readers to understand the availability of data underlying the research results described in the article. The statement may refer to original data generated in the course of the study or to third-party data analysed in the article. The statement should describe and provide means of access, where possible, by linking to the data or providing the required accession numbers for the relevant databases or DOIs.



\bibliographystyle{mnras}
\bibliography{tno2} % if your bibtex file is called example.bib

%\begin{thebibliography}{99}
%\end{thebibliography}

%%%%%%%%%%%%%%%%%%%%%%%%%%%%%%%%%%%%%%%%%%%%%%%%%%

%%%%%%%%%%%%%%%%% APPENDICES %%%%%%%%%%%%%%%%%%%%%

%\appendix


%If you want to present additional material which would interrupt the flow of the main paper, it can be placed in an Appendix which appears after the list of references.

%%%%%%%%%%%%%%%%%%%%%%%%%%%%%%%%%%%%%%%%%%%%%%%%%%

% Don't change these lines
\bsp % typesetting comment
\label{lastpage}
\end{document}